\pdfoutput=1

\documentclass[11pt]{article}

\usepackage[final]{acl}

\usepackage{times}
\usepackage{latexsym}

\usepackage[T1]{fontenc}

\usepackage[utf8]{inputenc}

\usepackage{microtype}

\usepackage{inconsolata}
\usepackage{pifont}
\usepackage{graphicx}

\usepackage{listings}
\usepackage{markdown}
\definecolor{mygreen}{rgb}{0,0.6,0}
\definecolor{mymauve}{rgb}{0.58,0,0.82}
\usepackage{xspace}
\usepackage{xcolor}
\usepackage{booktabs}
\usepackage{nicematrix}
\usepackage{cleveref}
\usepackage{adjustbox}
\usepackage{listings-rust}
\usepackage{caption}
\usepackage{subcaption}

\newcommand{\benchname}{\textsc{PseudoEval}\xspace}
\newcommand{\livecb}{LiveCodeBench\xspace}

\newcommand{\smalltitle}[1]{\smallskip\noindent\textbf{#1.}\xspace}
\newcommand{\codef}[1]{\texttt{#1}\xspace}
\newcommand{\eg}{e.g.}
\newcommand{\ie}{i.e.}

\newcommand{\passk}{Pass@k\xspace}

\newcommand{\dsr}{DeepSeek-R1\xspace}

\newcommand{\name}{\textsc{PseudoEval}\xspace}

\title{Isolating Language-Coding from Problem-Solving: Benchmarking LLMs with PseudoEval}

\author{Jiarong Wu\textsuperscript{1} \\  
  \texttt{jwubf@connect.ust.hk} \\\And
  Songqiang Chen\textsuperscript{1} \\
  \texttt{i9s.chen@connect.ust.hk} \\\And
  Jialun Cao\textsuperscript{1,*} \\
  \texttt{jcaoap@cse.ust.hk} \\
  \AND
  {\bf Hau Ching Lo\textsuperscript{1}} \\
  \texttt{hcloaf@connect.ust.hk} \\\And
  {\bf Shing-Chi Cheung\textsuperscript{1,*}} \\
  \texttt{scc@cse.ust.hk} \\
  \AND
  \textnormal{
  \text{\textsuperscript{1}The Hong Kong University of Science and Technology,
  \textsuperscript{*}Corresponding Authors}}
  }

\begin{document}
\maketitle
\begin{abstract}
Existing code generation benchmarks for Large Language Models (LLMs) such as HumanEval and MBPP are designed to study LLMs' end-to-end performance, where the benchmarks feed a problem description in nature language as input and examine the generated code in specific programming languages. However, the evaluation scores revealed in this way provide a little hint as to the bottleneck of the code generation -- whether LLMs are struggling with their problem-solving capability or language-coding capability.
To answer this question, we construct \name, a multilingual code generation benchmark that provides a solution written in pseudocode as input.
By doing so, the bottleneck of code generation in various programming languages could be isolated and identified. Our study yields several interesting findings. For example, we identify that the bottleneck of LLMs in Python programming is problem-solving, while Rust is struggling relatively more in language-coding.
Also, our study indicates that problem-solving capability may transfer across programming languages, while language-coding needs more language-specific effort, especially for undertrained programming languages.
Finally, we release the pipeline of constructing \name to facilitate the extension to existing benchmarks. \name is available at: \url{https://anonymous.4open.science/r/PseudocodeACL25-7B74/}.
\end{abstract}

\section{Introduction}\label{sec:introduction}

\begin{figure*}[t]
    \centering
    \includegraphics[width=0.97\linewidth]{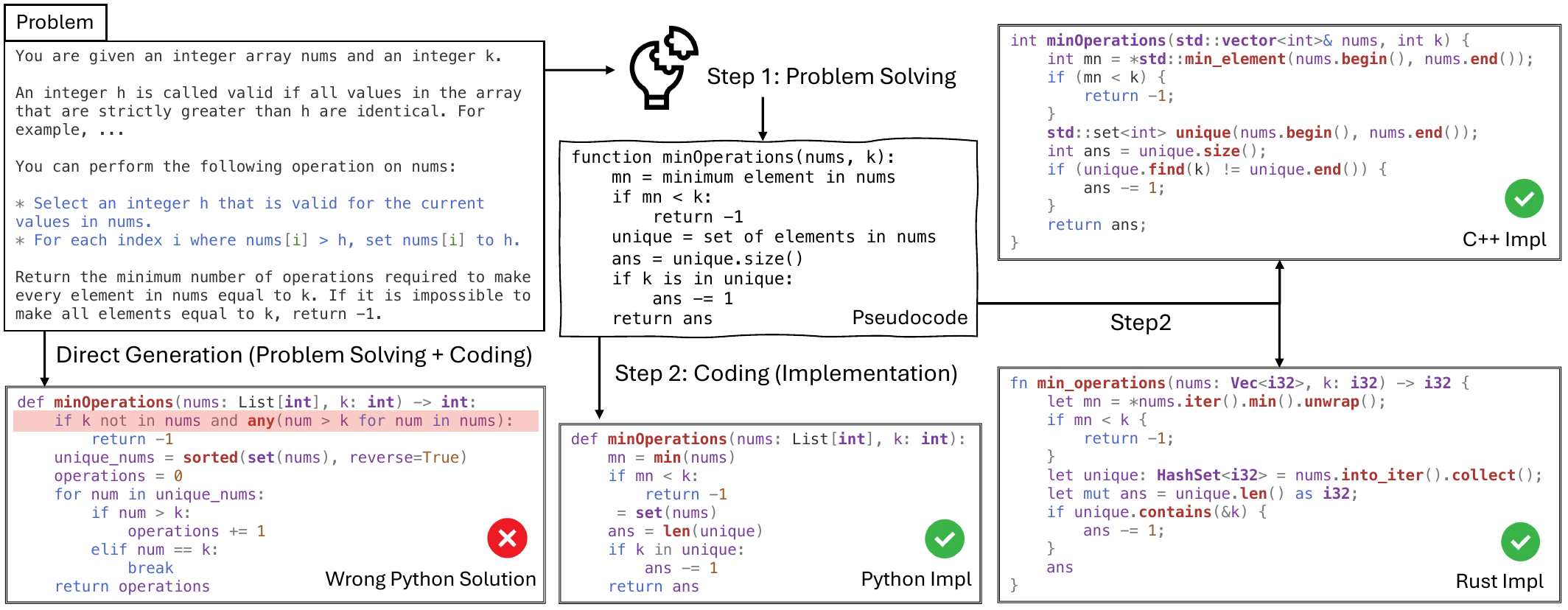}
    \caption{Motivating example}
    \label{fig:motivating-example}
\end{figure*}

Large Language Models (LLMs) have exhibited impressive proficiency in aiding software development, particularly in the realm of code generation. Existing code generation benchmarks, such as HumanEval~\cite{humaneval}, typically present a natural language description (e.g., ``return a list with elements incremented by 1'') and require LLMs to generate code that fulfills the described functionality. On the HumanEval leaderboard, various LLMs have achieved scores close to perfection (at most 99.4\%~\footnote{Result on Feb 13,2024, from \url{https://paperswithcode.com/sota/code-generation-on-humaneval}}). However, on another benchmark known for minimal contamination, LiveCodeBench~\cite{livecb}, the highest score recorded is 76.5\%~\footnote{Result on Feb 13, 2024, from \url{https://livecodebench.github.io/leaderboard.html}}. The highest score drops to 52.2\% for problems in the hard category.

However, what do these scores truly imply? When scores approach perfection, does it genuinely imply that the LLMs have nearly attained the capability to replace Python developers? The answer seems to be \textit{no}. Numerous studies have revealed significant shortcomings in LLMs' code generation capabilities, such as producing code with syntactic errors, code that does not meet the intended requirements, or code with low-level implementation mistakes. Yet, merely summarizing these phenomena as ``hallucinations''~\cite{li2023halueval,xu2024hallucination,zhang2023siren,dhuliawala2023chain, issta25llmhal} is an oversimplification. We seek to understand \textbf{\textit{what the bottleneck of code generation is}} -- Is it due to a lack of \textbf{\textit{problem-solving capability}} or \textbf{\textit{language-coding capability}}, or both?

To facilitate this study, we constructed a multi-lingual code generation benchmark, \name, with 1,060 subjects with not only problem-solution pairs but also intermediate solutions represented as pseudocode, which serve to isolate the problem-solving capability from the language-coding capability.
Take Figure~\ref{fig:motivating-example} for example. Given a problem description (upper-left corner), the existing end-to-end code generation benchmarks typically examine whether the generated code (lower-left corner) is implemented correctly and report a binary result (pass or fail) as the evaluation output.
However, the binary result gives little hint of the bottleneck, i.e., it is still unclear \textbf{\textit{whether LLMs are incapable of coming up with solutions for this problem}} or \textbf{\textit{\textbf{suffering from language-specific implementation}}} such as writing syntactic- or semantic-correct code in certain programming languages such as C++ or Rust. With \name, the assessment would yield clearer results -- by breaking the end-to-end task down into two steps.
One could observe when providing the solution (Pseudocode in the middle of Figure~\ref{fig:motivating-example}), LLMs can successfully code it in three languages (Python, C++, and Rust),
while all experimental LLMs failed to solve this easy-tagged problem without the provided solution,
\textbf{\textit{indicating the bottleneck for this problem is more on the problem-solving than language-coding capability}}. Furthermore, to expand the usefulness of \name, we explore four research questions (RQs). 

\textbf{\textit{RQ1. To what extent can the provided pseudocode improve the correctness of code generation?}} This RQ provides an overall profiling of the performance from the question description and the pseudocode. Understanding performance differences with and without pseudocode across different LLMs/programming languages/difficulties of the questions helps identify \textit{the bottleneck of code generation in different programming languages}. 

\textit{\textbf{RQ2. To what extent can the solution from one programming language benefit the code generation in another programming language?}} This RQ extends the study from monolingual to multilingual observation. It explores whether the pseudocode derived from codes in one programming language could benefit the code generation in another programming language. The results could \textit{give hints of the possibility of transferring problem-solving capability across programming languages.}

\textit{\textbf{RQ3. Can different inference strategies yield significantly different observations?}} Different promptings and attempts may yield different performances. This RQ explores whether the observation of the bottleneck (problem-solving or implementation) would significantly vary under different inference strategies.

\textbf{\textit{RQ4. What is the difference between human-written pseudocode and auto-extracted pseudocode?}} The pseudocode in \name is automatically extracted from the solution code. However, no prior study has been made to examine the quality of the pseudocode quantitatively. In this RQ, we compare the difference between the human-written pseudocode and the auto-extracted pseudocode in terms of token lengths, lines of code, and the LLMs' performance with pseudocode generated in both ways. The study provides more evidence to demonstrate the quality of the pseudocode in \name, and once assured, the auto-extraction we proposed could facilitate the extension to existing benchmarks.

Our study yields interesting observations. First, \textbf{\textit{the bottleneck}} in Python code generation is problem-solving, while C++ and Rust struggle relatively more in language-coding. Second, most solutions are \textbf{\textit{language-agnostic}}, indicating it may be enough for LLMs to learn problem-solving skills in certain programming languages and put more effort into the coding capability in programming languages. Third, the auto-generated pseudocode is \textbf{\textit{comparable or even better}} quality than human-written ones. Thus, it is feasible to extend the existing benchmarks with pseudocode with our pipeline. The contribution of this paper includes:

{\footnotesize \ding{108}} \textbf{\textit{Problem Decomposition}}: We break down the end-to-end code generation (from problem description to implementation) into a two-step evaluation (from problem description in natural language or from solutions in pseudocode). By doing so, the bottleneck of code generation in various programming languages could be isolated and identified.

{\footnotesize \ding{108}} \textit{\textbf{Benchmark \name}}: We constructed a multi-lingual (Python, C++, and Rust) code generation benchmark with 1,060 subjects comprising not only problem description in natural language and corresponding tests, but also the intermediate solutions in the form of pseudocode. The benchmark enables exploration of the bottleneck in code generation, provides clear criteria for pseudocode construction, and makes available a pipeline that implements a workflow automating the construction process. With it, one could refurbish existing code generation benchmarks easily. 

{\footnotesize \ding{108}} \textit{\textbf{Insight}}: We isolate LLMs' capabilities for code generation into problem-solving and language-coding. Our study finds that the bottleneck of generating code in different programming languages is different. Our study further suggests that problem-solving capability may transfer across programming languages while the coding capability for programming languages beyond the most popular ones remains to be improved.

\section{Problem Definition}
\subsection{Task Definition}

As shown in \Cref{fig:motivating-example}, the problem of LLM code generation comprises two tasks.

\emph{(1) Problem Solving,} which analyzes the problem and reasons a \emph{solution} as output.
The granularity of a solution can vary from a one-sentence description of the core algorithm to a pseudocode with a clear control flow and data manipulation.%
 
\emph{(2) Language Coding,} which transforms the solution into a piece of compilable and executable code that implements the key logic and data manipulation in a target programming language.

The two tasks exercise distinct abilities of LLMs, and previous code generation benchmarks such as LiveCodeBench~\cite{livecb} evaluate them inseparably.
This paper studies the coding ability of LLMs by isolating it from their problem-solving ability using pseudocode. %

\subsection{Pseudocode Criteria}\label{subsec:criteria}
Although there is no universal concrete standard for pseudocode, conventions such as guidebooks have been commonly adopted. %
To study the coding ability of LLMs, we adopt a set of criteria to prepare the pseudocode for \benchname. The criteria are designed based on the features of pseudocode in textbooks, guidebooks, and research papers. %

\smalltitle{Completeness}
The pseudocode should be mapped to a piece of implementation code \emph{without ambiguity}, \eg, a competent programmer should be able to implement the pseudocode solving the given problem in a specific programming language.
If given a piece of implementation code, one can obtain a trivial but complete pseudocode by line-by-line code translation~\cite{spoc}.

\smalltitle{Language-agnostic}
The pseudocode should describe a language-agnostic solution. It should not be tied to specific language features, such as the \codef{yield} expression in Python and the pointer manipulations in C++.
In particular, explicit type information (\eg, \codef{vector} in C++) and type conversion should be absent.
The language-agnostic criterion facilitates a fair evaluation of LLMs' coding abilities in different target programming languages with the same pseudocode.

\smalltitle{Conciseness}
A pseudocode should be concise, which can be measured by the lines of code and the number of tokens.
In practice, software developers tend to sketch solutions concisely.
Also, verbose pseudocode with implementation details may not help differentiate the abilities of stronger models and weaker models.
An interesting case (\Cref{subsec:simplication}) in our study is to simplify a well-known algorithm, Sieve of Eratosthenes, and customize its use in pseudocode.
LLMs with higher coding capability can successfully implement the pseudocode, while weaker LLMs have lower success rates and even drop to zero when the target language is Rust.

Following the above criteria, we define more specific rules (\Cref{sec:prompts}) to prompt DeepSeek-R1 to convert an implementation code into a pseudocode to construct \benchname (\Cref{sec:benchmark-construction}).

\section{Dataset Construction}\label{sec:benchmark-construction}
\begin{figure*}
    \centering
    \includegraphics[width=0.82\linewidth]{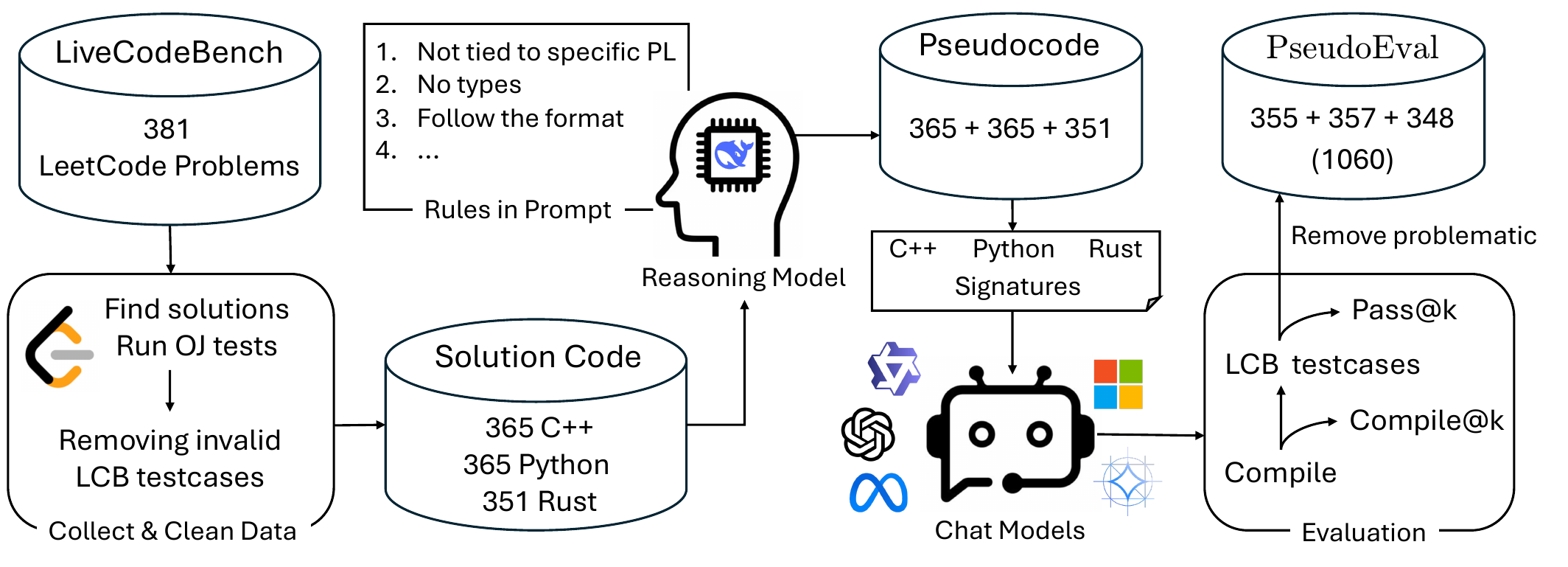}
    \caption{Workflow of constructing the \benchname dataset and empirical study }
    \label{fig:workflow}
\end{figure*}

To build \benchname, we design a pipeline implementing an automated workflow in \Cref{fig:workflow} to collect user-submitted solutions on LeetCode and distill pseudocode solutions from them using a recent reasoning model DeepSeek-R1. The pipeline may also be adapted to refurbish other existing code generation benchmarks.

\smalltitle{Data Source}
To lessen the data leakage threat, we select user-submitted solutions based on the problems most recently collected by \livecb~\cite{livecb}. These are the latest programming problems released after the training cut-off dates of popular LLMs.
In other words, we select the most recent subset of problems indexed by \livecb at LeetCode. We further collect the 
corresponding user-submitted solutions from LeetCode.
For each problem, we manually collect the most popularly voted solutions in Python, C++, and Rust, respectively.

\smalltitle{Task Cleaning}
To ensure the correctness of the collected user-submitted solutions,
we run each solution via the LeetCode online judge to ensure the solution passes all mandated tests.
If the most popularly voted solutions fail (usually due to the update of problems/tests), we collect another solution that passes the updated tests.
The study of our research questions requires evaluating the correctness of many generated codes. Submitting all of them to the LeetCode online judge for correctness validation is inappropriate. Therefore, we collect the published tests deduced by \livecb and use them to evaluate the correctness of the generated codes in our study. However, these \livecb tests are deduced by LLMs and subject to noises. %
We consider a deduced \livecb test noisy if it fails the collected solutions. %
In total, we find 16 noisy instances and exclude them from our study. 
After cleaning, we collect 365 solutions in C++ and Python and 351 solutions in Rust.

\smalltitle{Code to Pseudocode}
Each pseudocode used to evaluate the coding capability of LLMs is generated by the reasoning model \dsr~\cite{ds-r1} given a solution code and a detailed list of rules (\Cref{sec:prompts}) that the output pseudocode needs to satisfy, i.e., the criteria in \Cref{subsec:criteria}.
For example, the pseudocode should not contain explicit types like 32-bit or 64-bit integers and language-specific operations like \codef{yield} in Python.

We choose a reasoning model over a chat model like GPT-4o. Our pilot experiments find that chat models often fail to obey the rules in a long context or just write the pseudocode line by line without undergoing a substantial thinking process.
The prompt we use consists of only the user query without a system message or few-shot examples, as suggested by the DeepSeek team~\cite{ds-r1}.
We also follow their experiment setting (\codef{temperature=0.6}, \codef{top\_p=0.95}).
One pseudocode sample is obtained for each selected user-submitted solution due to the limited access to the R1 service and the incurred time latency.

\smalltitle{Pseudocode Quality Assessment}
{To remove incorrect R1-generated pseudocode, we use LLMs to generate code from the R1-generated pseudocode using our study setup and remove the tasks where \emph{NO LLMs} can pass the task with ten attempts.
Finally, we remove 22 subjects where R1 hallucinates a pseudocode with incorrect logic (e.g., adding an incorrect condition), and keep 1,059 subjects.
Besides, we compare the lengths and effectiveness of pseudocode annotated by R1 and humans for randomly sampled subjects in RQ4. The results also suggest good quality of the retained pseudocode.} %

\section{Experiment}\label{sec:experiment}

\smalltitle{Data}
To facilitate the comparison across programming languages, 999 (333 $\times$ 3) experiment subjects are drawn from the intersected programming tasks for C++ (355), Python (357), and Rust (348).

\smalltitle{Metrics}
The correctness of the generated programs is calculated by their \passk rates on the tests published by \livecb. %
The conciseness of pseudocode is measured by their lengths regarding the number of Byte-Pair Encoding (BPE) tokens and lines of codes.

\begin{table*}[t]
    \centering
    \definecolor{c1}{HTML}{FAFDFD}
    \begin{adjustbox}{max width=\textwidth}
    \begin{NiceTabular}{c || ccc|ccc || ccc|ccc || ccc|ccc }
    \CodeBefore
    \Body
        \toprule
         \Block{3-1}{} & \Block{1-6}{Python} &&&&& & \Block{1-6}{C++} &&&&& & \Block{1-6}{Rust} &&&&&  \\
         \cmidrule{2-19}
         & \Block{1-3}{from Problem} &&& \Block{1-3}{from Pseudocode} &&& \Block{1-3}{from Problem} &&& \Block{1-3}{from Pseudocode} &&& \Block{1-3}{from Problem} &&& \Block{1-3}{from Pseudocode} &&\\
         & Easy & Med & Hard & Easy & Med & Hard & Easy & Med & Hard & Easy & Med & Hard & Easy & Med & Hard & Easy & Med & Hard  \\ 

\midrule
GPT-4o-mini & 0.82 & 0.32 & 0.07 & 0.95 & 0.88 & 0.76 & 0.78 & 0.27 & 0.13 & 0.91 & 0.81 & 0.67 & 0.70 & 0.21 & 0.07 & 0.84 & 0.57 & 0.37  \\
Qwen32B     & 0.90 & 0.56 & 0.20 & 0.93 & 0.94 & 0.79 & 0.87 & 0.50 & 0.20 & 0.92 & 0.86 & 0.67 & 0.81 & 0.47 & 0.17 & 0.91 & 0.69 & 0.60  \\
Qwen32Bq4   & 0.87 & 0.56 & 0.22 & 0.92 & 0.94 & 0.79 & 0.85 & 0.52 & 0.20 & 0.91 & 0.86 & 0.70 & 0.80 & 0.44 & 0.14 & 0.90 & 0.70 & 0.57  \\
Qwen14B     & 0.80 & 0.51 & 0.14 & 0.97 & 0.90 & 0.70 & 0.78 & 0.50 & 0.10 & 0.95 & 0.82 & 0.56 & 0.73 & 0.35 & 0.05 & 0.88 & 0.55 & 0.35  \\
Qwen7B      & 0.68 & 0.34 & 0.11 & 0.86 & 0.81 & 0.53 & 0.69 & 0.36 & 0.10 & 0.86 & 0.72 & 0.39 & 0.55 & 0.23 & 0.01 & 0.71 & 0.43 & 0.15  \\
Gemma9B     & 0.49 & 0.09 & 0.04 & 0.84 & 0.68 & 0.45 & 0.44 & 0.10 & 0.06 & 0.84 & 0.55 & 0.26 & 0.30 & 0.03 & 0.01 & 0.41 & 0.23 & 0.08  \\
Llama3-8B   & 0.40 & 0.08 & 0.01 & 0.68 & 0.60 & 0.47 & 0.32 & 0.07 & 0.03 & 0.68 & 0.48 & 0.25 & 0.25 & 0.03 & 0.01 & 0.53 & 0.25 & 0.09  \\
Llama3-3B   & 0.27 & 0.04 & 0.00 & 0.54 & 0.41 & 0.23 & 0.20 & 0.03 & 0.01 & 0.46 & 0.26 & 0.15 & 0.15 & 0.00 & 0.00 & 0.33 & 0.07 & 0.01  \\
Phi4-14B    & 0.66 & 0.28 & 0.04 & 0.92 & 0.83 & 0.77 & 0.66 & 0.27 & 0.10 & 0.88 & 0.74 & 0.43 & 0.61 & 0.18 & 0.05 & 0.77 & 0.49 & 0.27  \\
Phi3.5-4B   & 0.44 & 0.08 & 0.03 & 0.67 & 0.46 & 0.30 & 0.36 & 0.05 & 0.03 & 0.54 & 0.29 & 0.09 & 0.14 & 0.00 & 0.02 & 0.22 & 0.04 & 0.00  \\
        
\midrule
\Block{2-1}{\textit{Average}}
& 0.63 & 0.29 & 0.09 & 0.83 & 0.75 & 0.58 & 0.60 & 0.27 & 0.10 & 0.80 & 0.64 & 0.42 & 0.50 & 0.19 & 0.05 & 0.65 & 0.40 & 0.25 \\
	& 	& 	& 	& \textit{\footnotesize31\%$\uparrow$} 	& \textit{\footnotesize160\%$\uparrow$} 	& \textit{\footnotesize573\%$\uparrow$} 	& 	& 	& 	& \textit{\footnotesize34\%$\uparrow$} 	& \textit{\footnotesize139\%$\uparrow$} 	& \textit{\footnotesize334\%$\uparrow$} 	& 	& 	& 	& \textit{\footnotesize29\%$\uparrow$} 	& \textit{\footnotesize107\%$\uparrow$} 	& \textit{\footnotesize370\%$\uparrow$} \\ \midrule
\textit{Overall} & \Block{1-3}{0.38} &&& \Block{1-3}{0.75 \textit{ \footnotesize(99\%$\uparrow$)}} &&& \Block{1-3}{0.35} &&& \Block{1-3}{0.66 \textit{ \footnotesize(87\%$\uparrow$)}} &&& \Block{1-3}{0.28} &&& \Block{1-3}{0.46 \textit{ \footnotesize(67\%$\uparrow$)}} & \\
\bottomrule

    \end{NiceTabular}
    \end{adjustbox}
    \caption{Pass@1 of generations from problem descriptions and pseudocode for easy, medium, and hard tasks}%
    \label{tab:overall-result}
\end{table*}

\smalltitle{Studied LLMs}
We study the code generation performance of ten diverse popular LLMs, including Qwen-series (Qwen-2.5-Coder 7B, 14B, 32B, 32B-Int(q)4) \cite{qwen2.5-coder}, Gemma-series (Gemma-2-9b) \cite{google-gemma-2}, Llama-series (Llama-3.1-8B and -3.2-3B) \cite{meta-llama-3-2}, Phi-series (Phi-4-14B and -3.5-4B) \cite{microsoft-phi-4}, and GPT-series (GPT-4o-mini) \cite{gpt-4o-mini-report}. Relatively more LLMs evaluated are under 15B parameters. This is to understand the {language-coding ability} of lighter, more deployable models.%

\smalltitle{Parameters}
We follow the setting suggested by LiveCodeBench to sample ten times of generations for each problem using a temperature of $0.2$ and top\_p of $0.95$.
We compare one-shot and zero-shot prompts for pseudocode-based code generation.

\smalltitle{Experiment Environment}
The experiments are conducted on a Linux server with two NVIDIA RTX 6000Ada GPUs. The commercial GPT-4o-mini and the primary DeepSeek-R1 are accessed via API calls. Other open-weight LLMs are deployed locally on the server with the vLLM engine.

\subsection{RQ1: Overall Performance}
\begin{table}[t]
    \centering
    \begin{adjustbox}{max width=0.48\textwidth}
    \setlength{\tabcolsep}{4pt}
    \begin{NiceTabular}{c | ccc | ccc | ccc}
    \CodeBefore
    \Body
        \toprule
         \Block{2-1}{}
         & $P_\textrm{Py}$ & $P_\textrm{C++}$ & $P_\textrm{Rust}$ & $P_\textrm{Py}$ & $P_\textrm{C++}$ & $P_\textrm{Rust}$ & $P_\textrm{Py}$ & $P_\textrm{C++}$ & $P_\textrm{Rust}$ \\
         & \Block{1-3}{$\rightarrow$ {{Python Code}}} && & \Block{1-3}{$\rightarrow$ {{C++ Code}}} && & \Block{1-3}{$\rightarrow$ {{Rust Code}}} &&  \\
         \midrule

GPT-4o-mini & 0.89 & 0.90 & \textbf{0.91} & 0.80 & \textbf{0.82} & \textbf{0.82} & 0.61 & \textbf{0.64} & 0.63  \\ 
Qwen32B &     \textbf{0.91} & 0.90 & 0.90 & 0.81 & \textbf{0.85} & 0.84 & 0.70 & 0.74 & \textbf{0.75}  \\ 
Qwen32Bq4 &   \textbf{0.91} & 0.90 & 0.90 & 0.81 & \textbf{0.86} & 0.85 & 0.70 & 0.73 & \textbf{0.75}  \\ 
Qwen14B &     \textbf{0.89} & 0.88 & 0.88 & 0.76 & \textbf{0.82} & 0.80 & 0.59 & \textbf{0.63} & \textbf{0.63}  \\ 
Qwen7B &      0.78 & \textbf{0.83} & 0.81 & 0.65 & \textbf{0.72} & 0.68 & 0.45 & \textbf{0.50} & 0.48  \\ 
Gemma9B &     0.69 & \textbf{0.73} & 0.69 & 0.56 & \textbf{0.60} & 0.56 & 0.27 & \textbf{0.29} & 0.27  \\ 
Llama3-8B &   0.61 & \textbf{0.65} & 0.60 & 0.46 & \textbf{0.51} & \textbf{0.51} & 0.28 & 0.30 & \textbf{0.32}  \\ 
Llama3-3B &	  0.42 & \textbf{0.44} & \textbf{0.44} & 0.28 & \textbf{0.31} & 0.30 & \textbf{0.17} & 0.14 & 0.15  \\ 
Phi4-14B &	  \textbf{0.85} & 0.83 & \textbf{0.85} & 0.72 & \textbf{0.74} & 0.73 & 0.53 & \textbf{0.56} & 0.55  \\ 
Phi3.5-4B & \textbf{0.51} & \textbf{0.51} & 0.47 & 0.31 & \textbf{0.35} & 0.30 & 0.11 & \textbf{0.12} & 0.10 \\ 
\midrule
\textit{Average}
& 0.75 & \textbf{0.76} & 0.75 & 0.62 & \textbf{0.66} & 0.64 & 0.44 & \textbf{0.47} & 0.46 \\ 
    \bottomrule
    \end{NiceTabular}
    \end{adjustbox}
    \caption{Pass@1 of code generation with pseudocode derived from different programming languages}
    \label{tab:compare-lang}
\end{table}

To understand the {language-coding capability} of LLMs, we analyze the quality of the codes that LLMs generated from pseudocode. 
Each column in \Cref{tab:overall-result} presents the Pass@1 rate of LLMs in generating programs of a specified programming language based on the pseudocode derived from the solution codes in the same language.
We also list LLMs' Pass@1 rates when directly generating codes from problem descriptions using the prompt adopted by LiveCodeBench as a reference.

\smalltitle{Effect of Pseudocode} 
All ten LLMs achieve significantly higher Pass@1 rates on all programming languages when generating programs from pseudocode than from problem descriptions. 
Specifically, the overall Pass@1 rates on all difficulties increase from 0.38, 0.35, 0.28 to 0.75, 0.66, 0.46 on Python, C++, and Rust, respectively.
The results suggest that the solutions encoded in pseudocode help LLMs generate more correct programs.
As such, we consider that \textit{problem-solving ability is a key bottleneck common to LLMs}.
Regarding programming languages, all LLMs exhibit the largest performance gain in Python programming (+99\% on average), followed by C++ (+87\% on average). Performance improvement on Rust is the least (67\% on average) yet is still significant. As Rust coding is lower in resource availability \cite{humanevalx,cao2025should}, the result suggests \textit{the correlation between language-coding ability and the prevalence of the language in corpus}.

\smalltitle{Language-Coding Capability} 
LLMs' language-coding capability varies across programming languages. 
Given pseudocode, most LLMs can generate correct implementations in Python; while they still cannot generate correct Rust implementations for many tasks. For example, given pseudocode, the Python, C++, and Rust Pass@1 rates are 0.89, 0.82, 0.63 for GPT-4o-mini and 0.85, 0.74, 0.55 for Phi-4, respectively.
It suggests that \textit{the bottleneck of LLMs in code generation is problem-solving, while as to Rust and C++, they are struggling relatively more in language coding}. 

\smalltitle{Difficulty-Wise} LLMs show the most improvement in Pass@1 rates on the hard tasks (573\%$\uparrow$, 334\%$\uparrow$, and 370\%$\uparrow$ in Python, C++, and Rust, respectively), followed by the medium tasks (160\%$\uparrow$, 139\%$\uparrow$, and 107\%$\uparrow$) and then the easy tasks (31\%$\uparrow$, 34\%$\uparrow$, and 29\%$\uparrow$). Since hard tasks depend more on problem-solving ability, the result echos our conclusion that problem-solving is a key bottleneck.

\smalltitle{Model-Wise} 
Given pseudocode, the Python Pass@1 rates of the best-performing studied LLM, QWen32B (and its quantified variant QWen32Bq4), on easy, medium, and hard tasks significantly increase from around 0.90, 0.56, 0.20 to 0.93, 0.94, 0.79, respectively, followed by QWen14B (0.80, 0.51, 0.14 $\rightarrow$ 0.97, 0.90, 0.70) and GPT-4o-mini (0.82, 0.32, 0.07 $\rightarrow$ 0.95, 0.88, 0.76). Similar trends are observed in C++ and Rust.
This indicates such powerful LLMs likely have mature language-coding capabilities, particularly in Python.
Most of their bottleneck in solving LiveCodeBench tasks may reside in the problem-solving procedure. 
In comparison, although the smaller models show improvement in Pass@1 rates given pseudocode, their generations based on pseudocode are still error-prone.
For example, almost all Pass@1 rates on Llama-3.1-8B, Llama-3.2-3B, and Phi-3.5-4B for medium and hard tasks are still below 0.50. 
This suggests that \textit{regarding the ability to implement a given programming logic, smaller LLMs are much inferior to the larger LLMs}.

\smalltitle{Worsening Cases}
We noticed a few cases where providing pseudocode degrades the models' performance. 
For example, when writing Python code in 16 problems, %
GPT-4o-mini shows a lower Pass@1 when referring to pseudocode than solving problems directly.
Our analysis of the failure cases suggests that in some cases (e.g., \Cref{subsec:worsening}), LLM fails to understand expressions like ``cumulative sums'' of an array/list indicated in the pseudocode, which is expected to be implemented with \codef{accumulate()}. Such expressions may conversely mislead LLMs who were able to reason a correct solution from problem descriptions by themselves, particularly for easy problems.
In practice, ambiguous natural language expressions are inevitable in pseudocode or instructions, despite the semi-structured format of pseudocode. It is an interesting future work to detect and fix such noise.

In general, these interesting findings help us understand the language-coding ability of LLMs. They echo our motivation to gain a clearer picture of LLMs' coding capabilities by isolating the evaluation of their problem-solving and language-coding abilities by introducing pseudocode.

\subsection{RQ2: Cross-Programming Language}

Recalling that pseudocode abstracts language-specific details of solutions (\Cref{subsec:criteria}), we further investigate if pseudocode manifests generalizable effectiveness such that the pseudocode derived from solution code in a programming language can benefit LLMs in generating codes in another language.
\Cref{tab:compare-lang} presents the comparison. 

\smalltitle{Language-Wise} The pseudocode derived from solution codes written in any programming language ($P_{lang}$) effectively help LLMs gain much higher Pass@1 rates in comparison to the performance of generating from problems listed in \Cref{tab:overall-result} (e.g., 0.75{\textasciitilde}0.76 \textit{v.s.} 0.38 Pass@1 in Python generation). The result suggests that \textit{pseudocode can serve as a language-agnostic representation to hint LLMs about solution logic and guide LLMs generating programs in various programming languages}, which may shed light on cross-language tasks such as code translation and code search. 
Furthermore, we surprisingly found from the comparison among $P_{lang}$ that \textit{the pseudocode derived from C++ solutions help LLMs gain the highest Pass@1 in generating not only C++ but also Python and Rust programs on average}; meanwhile, pseudocode of Python solutions show inferior effectiveness for C++ and Rust generation.
Our manual analysis suggests the reason may be that Python codes often implement logic with various libraries, and thereby, the detailed idea to implement some features cannot be extracted into pseudocode. As a result, when there is no available corresponding library to use in C++ and Rust, the LLMs cannot correctly implement the logic. The analysis of the lengths of pseudocode derived from different programming languages also shows the trend as indicated in the 2nd{\textasciitilde}4th columns in \Cref{tab:loc-tokens}.%

\smalltitle{Model-Wise} The studied LLMs consistently gain improvement in Pass@1 rates with the help of pseudocode derived from any-language solution codes. 
Meanwhile, the most helpful programming language varies across LLMs. For example, Qwen14B and QWen32B work best when referring to the pseudocode derived from the solution code written in the target programming language, GPT-4o-mini prefers pseudocode of C++ or Rust solutions, and the others prefer C++.
This may indicate distinct LLMs need unique logic information implied in solution codes of specific programming languages.

\subsection{RQ3: Effects of Inference Strategies}
\begin{figure}
    \hspace{-8pt}\includegraphics[width=1.02\linewidth]{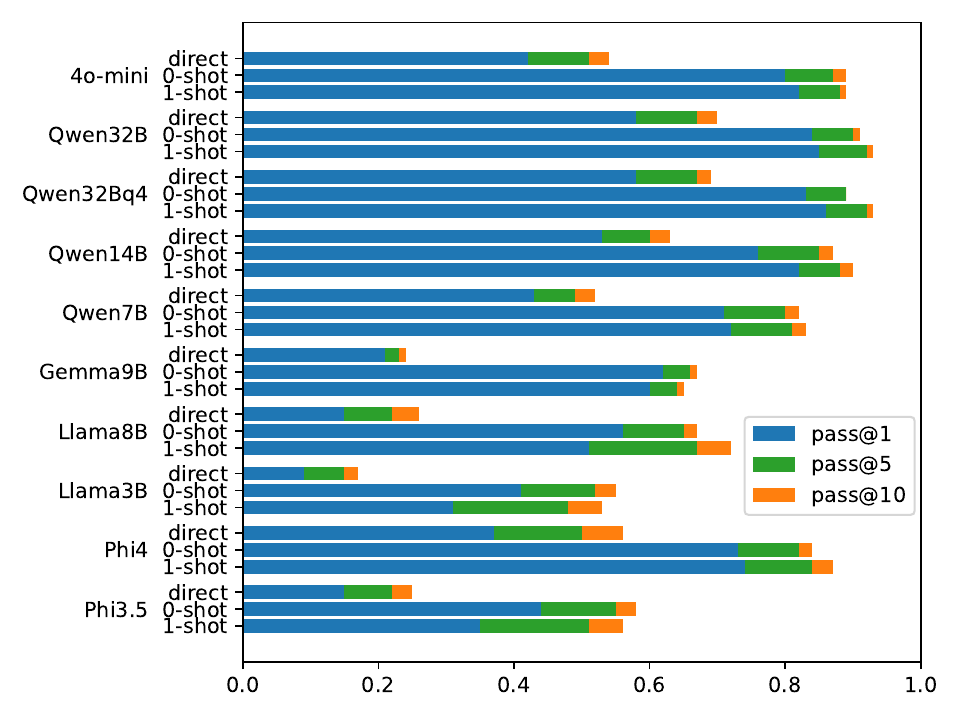}
    \vspace{-0.3cm}
    \caption{Zero-/one-shot Pass@\{1,5,10\} rates of C++ programs generated from pseudocode of C++ solutions}
    \label{fig:passk}
\end{figure}

We investigate if two typical configurations, i.e., whether using in-context learning (zero-shot $\rightarrow$ one-shot) and increasing the attempts (1$\rightarrow$5$\rightarrow$10), help improve the performance. 
\Cref{fig:passk} presents the Pass@\{1, 5, 10\} rates of LLMs when using zero-/one-shot prompts on generating C++ codes based on the pseudocode derived from C++ solutions. The results of other languages show consistent conclusions and are available at \Cref{sec:add-exp-results}.

\smalltitle{Zero-shot \textit{v.s.} One-shot}%
One-shot prompting benefits most LLMs but may disturb poorer LLMs like Phi-3.5-4B and Llama-3.1-3B which may not effectively handle long contexts. The result suggests using one-shot prompting to guide most LLMs better while driving smaller LLMs with more concise prompts. For consistency, in RQ1 and RQ2, we use one-shot prompts as a general setup for all LLMs.

\smalltitle{Pass@\{1, 5, 10\}} Increasing attempts also brings more chances of generating correct codes in the setup of pseudocode-based code generation, in particular for smaller LLMs like Llama-3.2-3B. For larger LLMs, 5 attempts may be appropriate considering cost-effectiveness.

\smalltitle{\textit{v.s.} Generating from Problem} (\textit{abbr.} direct) It is clear that Pass@1 rates of all LLMs when generating from pseudocode have already surpassed the effort of 10 attempts when generating from problem descriptions. The finding again echos our conclusion that \textit{problem-solving is the key bottleneck of current LLMs in code generation}. Besides, with pseudocode to hint the solution logic, \textit{one} attempt enables all LLMs except Phi-3.5 and Llama3B to outperform the Pass@{10} rates achieved by the commercial GPT-4o-mini generating from problems.

\subsection{RQ4: Automatically-generated v.s. Manually-written pseudocode}\label{subsec:resrq4}

\begin{table}[t]
    \centering
    \begin{adjustbox}{max width=\linewidth}
    \setlength{\tabcolsep}{4pt}
    \begin{NiceTabular}{c | c cc | cc }
    \CodeBefore
    \Body
        \toprule
        & \Block{1-3}{Source Code} &&& \Block{1-2}{Pseudocode} &\\
        & Lang & LoC & Tokens & LoC & Tokens \\
        \midrule
Manual &	C++ &	21.64 &	222.16 &	12.58	\textit{(-42\%)} &	151.45 	\textit{(-32\%)} \\
DeepSeek-V3 &	C++ &	21.64 &	222.16 &	18.84	\textit{(-13\%)} &	172.91 	\textit{(-23\%)} \\
DeepSeek-R1 &	C++ &	21.64 &	222.16 &	13.20	\textit{(-39\%)} &	122.31 	\textit{(-45\%)} \\
DeepSeek-R1 &	Rust &	18.45 &	219.33 &	12.71	\textit{(-31\%)} &	124.51 	\textit{(-43\%)} \\
DeepSeek-R1 &	Python &	15.47 &	156.89 &	11.93	\textit{(-23\%)} &	111.29 	\textit{(-29\%)} \\
    \bottomrule
    \end{NiceTabular}
    \end{adjustbox}
    \caption{Loc and tokens of a subset (55) of LCB tasks}
    \label{tab:loc-tokens}
\end{table}

\noindent We compare the pseudocode annotated by humans and DeepSeek-R1 on 55 sampled programming tasks. The manual annotation involved six developers with over five years of C++ coding experience; one pseudocode is annotated by an annotator and validated with another two. We mainly compare the simplicity and effectiveness of the two sets of pseudocode, which are essential qualities clear to measure based on lengths and pass rates.

\smalltitle{Simplicity} \Cref{tab:loc-tokens} lists the lengths of pseudocode annotated manually and automatically in the number of BPE tokens \cite{tiktoken} and lines of codes ignoring blank lines and comments. It is found that pseudocode, particularly the ones annotated by humans and DeepSeek-R1 (\textit{abbr.} R1), are much shorter than the source codes. This suggests that \textit{pseudocode is a simplified format to express the solution of programs}.
Besides, interestingly, although the manually annotated pseudocode include fewer lines than R1-generated ones, the manual ones are found to include more tokens than R1's. The reason is R1 tends to describe the logic more concisely.

\begin{table}[t]
    \centering
    \begin{adjustbox}{max width=0.48\textwidth}
    \setlength{\tabcolsep}{4pt}
    \begin{NiceTabular}{c | cccc | cccc }
    \CodeBefore
        \Body
        \toprule
        & \Block{1-4}{Manual Pseudocode} &&&& \Block{1-4}{DeepSeek-R1 Pseudocode} &&& \\
        & C@1 & P@1 & P@5 & P@10 & C@1 & P@1 & P@5 & P@10 \\
        \midrule
GPT-4o-mini & 0.97 & 0.60 & 0.72 & 0.75 & 0.94 & 0.78 & 0.83 & 0.84  \\ 
Qwen32B    & 1.00 & 0.65 & 0.71 & 0.71 & 0.98 & 0.81 & 0.87 & 0.87  \\ 
Qwen32Bq4  & 0.99 & 0.63 & 0.71 & 0.75 & 0.99 & 0.81 & 0.88 & 0.89  \\ 
Qwen14B    & 0.97 & 0.69 & 0.74 & 0.76 & 0.96 & 0.80 & 0.87 & 0.89  \\ 
Qwen7B     & 0.91 & 0.57 & 0.69 & 0.73 & 0.87 & 0.65 & 0.76 & 0.78  \\ 
Gemma9B    & 0.83 & 0.46 & 0.51 & 0.53 & 0.79 & 0.61 & 0.63 & 0.64  \\ 
Llama8B    & 0.83 & 0.47 & 0.62 & 0.65 & 0.79 & 0.48 & 0.59 & 0.64  \\ 
Llama3B    & 0.80 & 0.28 & 0.45 & 0.51 & 0.69 & 0.33 & 0.46 & 0.49  \\ 
Phi4       & 0.91 & 0.60 & 0.70 & 0.73 & 0.91 & 0.65 & 0.78 & 0.82  \\ 
Phi3.5     & 0.78 & 0.33 & 0.43 & 0.47 & 0.70 & 0.31 & 0.45 & 0.51  \\ 
\midrule
\textit{Average} & 0.90 & 0.53 & 0.63 & 0.66 & 0.86 & 0.62 & 0.71 & 0.74 \\ 
    \bottomrule
    \end{NiceTabular}
    \end{adjustbox}
    \caption{Compilation (C) and Pass (P) Rates of C++ code generation with pseudocode on 55 LCB tasks}
    \label{tab:passk-manual-r1}
\end{table}

\smalltitle{Effectiveness} We compare how effectively the automatically and manually annotated pseudocode guide LLMs in generating correct codes. As presented in \Cref{tab:passk-manual-r1}, the manual pseudocode help LLMs generate more compilable codes on the validation tasks. Meanwhile, surprisingly, \textit{the pseudocode generated by R1 contribute to higher Pass@k rates than the human-written ones}. The cause may be the gap between the expression style and knowledge preferences of humans and LLMs as reported by existing studies \cite{ase24-llmpreference}. %

Based on these detailed clues, we consider current SOTA reasoning LLMs like DeepSeek-R1 an effective helper to abstract reference codes into concise pseudocode with high accuracy.
They may offer feasible automation to abstract existing valuable code resources in GitHub or code generation benchmarks and facilitate studies on pseudocode.

\section{Related Work}\label{sec:related-work}

\smalltitle{Benchmarking End-to-End Code Generation}
Various benchmarks have been developed to assess LLMs in end-to-end code generation -- %
some benchmarks {\textit{broader the programming languages}} to evaluate. Classical benchmarks focus on Python programming \cite{humaneval, mbpp}; later, benchmarks considering other programming languages, e.g., Java \cite{JavaBench} and even multilingual \cite{humanevalx}, emerge. %
Some studies evaluate LLM programming {\textit{across different contexts}}, %
such as class-level \cite{classeval}, project-level \cite{deveval}, and repository-level \cite{repocoder, repoeval}, pushing the boundaries of LLM capabilities in real-world scenarios.
The performance of generating \textit{{code in different domains}} also attracts studies \cite{domaineval}. %
Several recent studies explored LLMs' code-generation capabilities {\textit{incorporating external techniques}}, for example, using RAG to retrieve codes \cite{ase24retrieval_repo, codesearchallyouneed_hu} and documents \cite{docragarxiv, docragccwan}, allowing LLMs to code with external resources.

Though these studies assess LLM's performance in various scenarios, they reveal relatively limited information about LLM's ability at steps within the end-to-end pipeline, e.g., coding a solution logic.

\smalltitle{Benchmarking Code Generation Using Pseudocode}
Only a few works have studied translating pseudocode into code. \citet{Dir17} propose a conceptual framework that breaks down pseudocode into XML elements. \citet{Kul19} explore potential mappings of pseudocode and C++ code using test cases. The SPoC dataset with 18K line-to-line mappings is built in the work. However, the fairly trivial line-by-line pseudocode may not accurately reflect the human-written pseudocode typically appearing in real-world software development. SPoC was later utilized by \citet{Ach22} to train two basic deep-learning models for pseudocode-to-code translation.
These studies worked on relatively small and trivial pseudocode snippets. They also barely compared the performance of code generators (in particular the advanced LLMs) or discussed the detailed abilities. %

\section{Insights from Study Results}

\indent\indent \ding{182} Code generation bottleneck differs across programming languages (PLs). %
One can improve end-to-end LLM programming performance for popular PLs like Python by boosting problem-solving abilities, whereas for less-trained languages like Rust, enhancing language-coding skills is crucial.

\ding{183} %
Problem-solving ability may transfer across PLs, which may allow LLMs' coding performance to be improved in a unified manner across PLs.

\ding{184} %
Reasoning models can effectively handle the code-to-pseudocode transformation. This enables easy creation of up-to-date benchmarks focusing on problem-solving capability, which may help relieve the current bottleneck and support cross-PL tasks.

These insights may shed light on enhancing LLMs in code generation and other cross-PL tasks, as well as guide human-LLM collaboration in the era of AI-driven low/zero-code development.

\section{Conclusion}\label{sec:conclusion}

To understand the bottlenecks in end-to-end code generation for LLMs, we introduce \name, a multilingual code generation benchmark incorporating pseudocode as input,
isolating the evaluation of language-coding from problem-solving capabilities. Empirical study results with \name reveal key insights about the bottlenecks identified for different programming languages, broad applicability of pseudocode across programming languages, and exceptional quality of automatically derived pseudocode. %

\clearpage

\section{Limitations}
\smalltitle{Pseudocode Samples}
Due to the limited access to DeepSeek-R1, the latency of response of reasoning models, and the costs of the subsequence inference, this study only sample one pseudocode for each problem.
As revealed in \Cref{subsec:resrq4}, a small portion of the generated pseudocode could be not semantic preserving and is filtered out from the final benchmark.
The thorough study on whether sampling multiple pseudocode or using a majority vote mechanism can further improve the pseudocode quality is left as future work.

\smalltitle{Problem Domain}
The current \name selects subjects from LiveCodeBench and their solutions on LeetCode, which are mainly algorithmic code for programming puzzles.
Although this meets the purpose of using pseudocode to present algorithms in practice, the daily software development scenarios such as implementing business logic are not covered.
It is unclear whether the performance gap between problem-to-code generation and pseudocode-to-code generation is also significant in such scenarios.
The future work to understanding this problem can be extending the workflow of \name to code generation benchmarks in different scenarios.

\smalltitle{Involved Programming Languages}
The programming languages studied in this paper are Python, C++, and Rust.
They represent three popular imperative programming languages, with a major difference in the type system.
Python is dynamic, C++ is static but weakly typed, and Rust is known for having a rigorous type checking mechanism. 
The results in RQ2 may shed light on similar languages such as Java, but may not apply to functional languages such as Haskell or low-resourced languages such as domain-specific languages.

\newpage

\bibliography{custom}

\appendix
\newpage

\section{Dataset}

\subsection{Legal Compliance and License}
The problems we use are from the LiveCodeBench, and the solutions we use to generate pseudocode are from LeetCode,
which are the publicly visible portions.
We did not include the user-submitted solutions in our final benchmark but their extracted pseudocode.
Following \citet{henry21} and LiveCodeBench~\cite{livecb}, we abide by Fair Use 107: ``the fair use of a copyrighted work, including such use by \ldots\xspace scholarship, or research, is not an infringement of copyright'', where fair use is determined by the purpose and character of the use, including whether such use is of a commercial nature or is for nonprofit educational purposes'', ``the amount and substantiality of the portion used in relation to the copyrighted work as a whole'', and ``the effect of the use upon the potential market for or value of the copyrighted work.''
The collected data in \name is used only for academic purposes.
Moreover, \name is used for benchmarking, and we do not use it for training models.

\subsection{Basic Stats}
\Cref{tab:loc-tokens-full} shows the number of files and statistics of source code and pseudocode from different languages in \name.
The statistics is basically consistent with the sampled subset in \Cref{subsec:resrq4}.
Each pseudocode corresponds to a problem in LiveCodeBench and can use its testcases to test the correctness of the code generated from the pseudocode.

\begin{table}[h]
    \centering
    \begin{adjustbox}{max width=\linewidth}
    \setlength{\tabcolsep}{4pt}
    \begin{NiceTabular}{c c | cc | cc  }
    \CodeBefore
    \Body
        \toprule
        \Block{2-1}{Language} & \Block{2-1}{\#Files} & \Block{1-2}{Source Code} && \Block{1-2}{Pseudocode} &  \\
        & & LoC & Tokens & LoC & Tokens \\
        \midrule
C++     & 355  & 18.20 & 192.23 & 13.05 & 122.58 \\
Rust    & 348  & 18.59 & 215.71 & 13.66 & 129.76 \\
Python	& 357  & 13.75 & 140.30 & 11.33 & 108.29 \\
    \bottomrule
    \end{NiceTabular}
    \end{adjustbox}
    \caption{Statistics of source code and pseudocode from different languages in \name }
    \label{tab:loc-tokens-full}
\end{table}

\section{Human Annotations}
Six programmers with more than five years of C++ experience participate in the annotation of pseudocode on 55 sampled C++ solution code.
Each annotated piece of pseudocode is validated by two other participants from the same group.

The approval from the ethics review board is exempted because the annotation procedure is not physically or mentally harmful and does not impose an intense workload in a short time.
The participants have been compensated according to the local legislation.
The consent to use the annotated data has been obtained from the participants.

\section{Case Study}

\subsection{Motivating Example}\label{subsec:motv-ex}
\Cref{lst:problem-motiv} lists the full problem and \Cref{lst:cpp-motiv} lists the user-submitted C++ solution where the pseudocode is converted from.
Note that the pseudocode simplifies the solution by replacing the map structure with a set structure.

\subsection{Simplifying Common Procedures}\label{subsec:simplication}
\Cref{lst:cpp-simp} shows a user-submitted C++ solution with detailed steps, and \Cref{lst:pseudo-simp} shows a concise but semantic-preserving pseudocode converted from the long solution.
Powerful LLMs such as GPT-4o-mini and Qwen32B can implement code correctly in all three languages, while smaller LLMs such as Phi-3.5 have lower success rates and even drop to zero when writing Rust codes. 

\subsection{Underflow in Rust}\label{subsec:rust-underflow}
\Cref{lst:rust-underflow} shows an example of a user-submitted Rust solution with the subtraction underflow problem.
Specifically, the variable \codef{pos} is from \codef{.len()} (line~5) and should be a \codef{usize} (unsigned) variable.
The user uses a nonstandard way to control the loop termination: when \codef{pos} is 0 and subtracts 1 from it in the release mode, it becomes the biggest unsigned integer, so the loop terminates because \codef{pos > arr.len()}.
However, such a coding style is not encouraged in Rust.
In the debug mode, the Rust program will panic (\ie, running into an invalid state because \codef{pos} is unsigned and should not underflow) and terminate the execution.

The pseudocode generated by DeepSeeek-R1 (\Cref{lst:pseudo-rust-underflow}) focuses on the solution logic, which does not contain such detailed type information and uses a more standard coding style (loop until \codef{pos} is negative).
Based on the pseudocode, only Qwen32B notices the possible sign problem and can generate code that correctly converts the type as \codef{isize} (line~6, \Cref{lst:rust-fix-underflow}), while all other less powerful models failed to do so.

\subsection{Worsening Pseudocode}\label{subsec:worsening}
\Cref{lst:python-acc}, \ref{lst:pseudo-python-acc}, and \ref{lst:python-acc-wrong-gen} show a case where the pseudocode generated from a Python solution misleads LLMs and causes a lower pass@1 compared with generating Python code from the problem.

\subsection{Failure of Pseudocode Generation}\label{subsec:fail-pseudogen}
\Cref{lst:python-being-wrongly-pseudogen} and \ref{lst:pseudo-wrong-condition} show a case of a Python solution and its generated pseudocode that is not semantic preserving.
The problem is at the last line, where the Python code will return \codef{max\_sum} if it is negative but not \codef{-inf}, while the pseudocode incorrectly assumes \codef{max\_sum} to be non-negative, possibly due to the hallucination problem in LLMs.

\section{Prompts}\label{sec:prompts}
\smalltitle{Generating Pseudocode}
\Cref{lst:pseudogen-prompt} is the prompt (a single user query as suggested by the DeepSeek team~\cite{ds-r1}) we use to query \dsr to generate pseudocode from Python code.
The prompts to generate pseudocode from C++ and Rust are similar with minor difference in the example code snippets.

\smalltitle{Generating Code from Pseudocode}
\Cref{lst:query-prompt-zero} is the zero-shot prompt, and \Cref{lst:query-prompt-one} is the one-shot prompt for generating Python code.
The prompts to generate C++ and Rust code are similar with language difference in the one-shot example.

\section{Additional Experiment Results}\label{sec:add-exp-results}
\begin{figure*}[t]
\centering
    \begin{subfigure}[b]{\textwidth}
         \centering
         \includegraphics[width=0.6\linewidth]{figs/compare_config_from_C++_to_C++.pdf} %
         \caption{To C++}
    \end{subfigure}
    \begin{subfigure}[b]{\textwidth}
         \centering
         \includegraphics[width=0.6\linewidth]{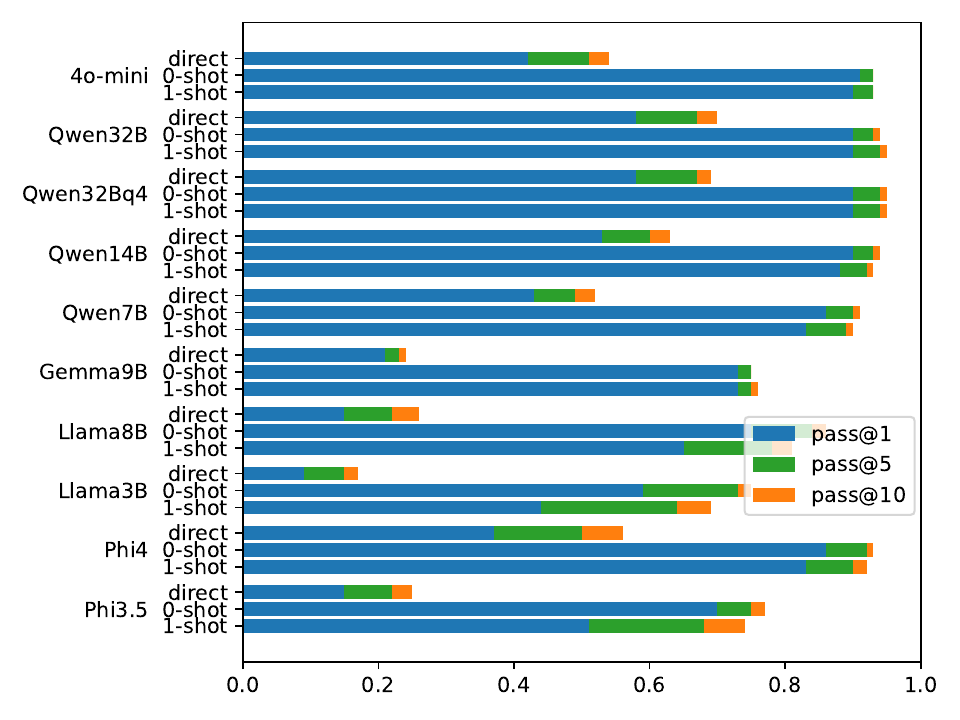}
         \caption{To Python}
    \end{subfigure}
    \begin{subfigure}[b]{\textwidth}
         \centering
         \includegraphics[width=0.6\linewidth]{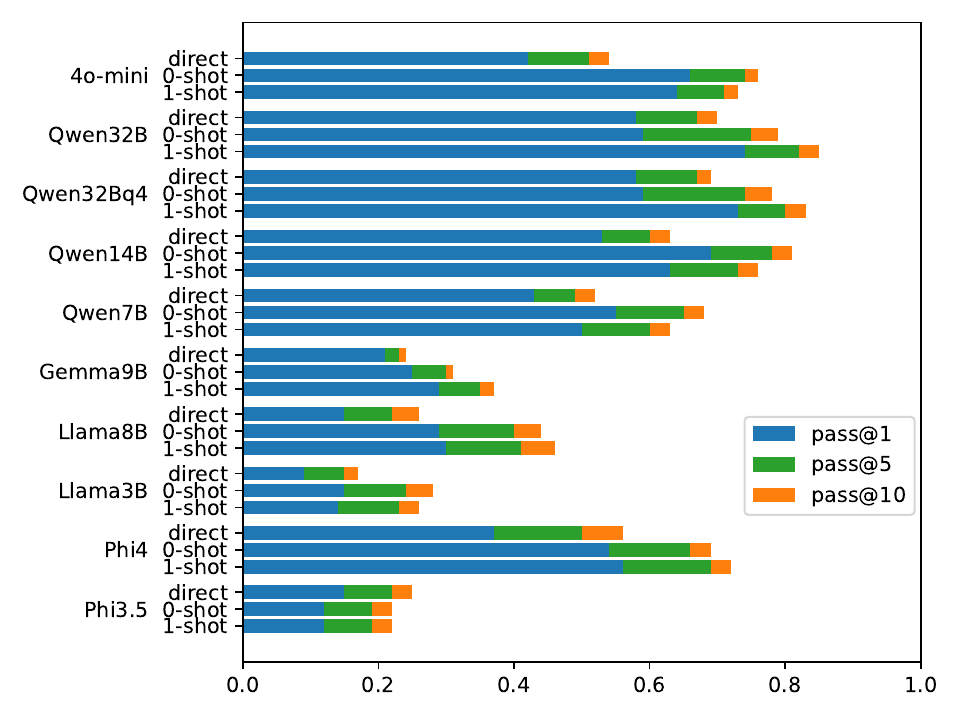}
         \caption{To Rust}
    \end{subfigure}
  \caption {Pass@k of code generation from pseudocode from C++ to all languages, compared with direct generation from problems}
  \label{fig:compare-config-cpp}
\end{figure*}

\begin{figure*}[t]
\centering
    \begin{subfigure}[b]{\textwidth}
         \centering
         \includegraphics[width=0.6\linewidth]{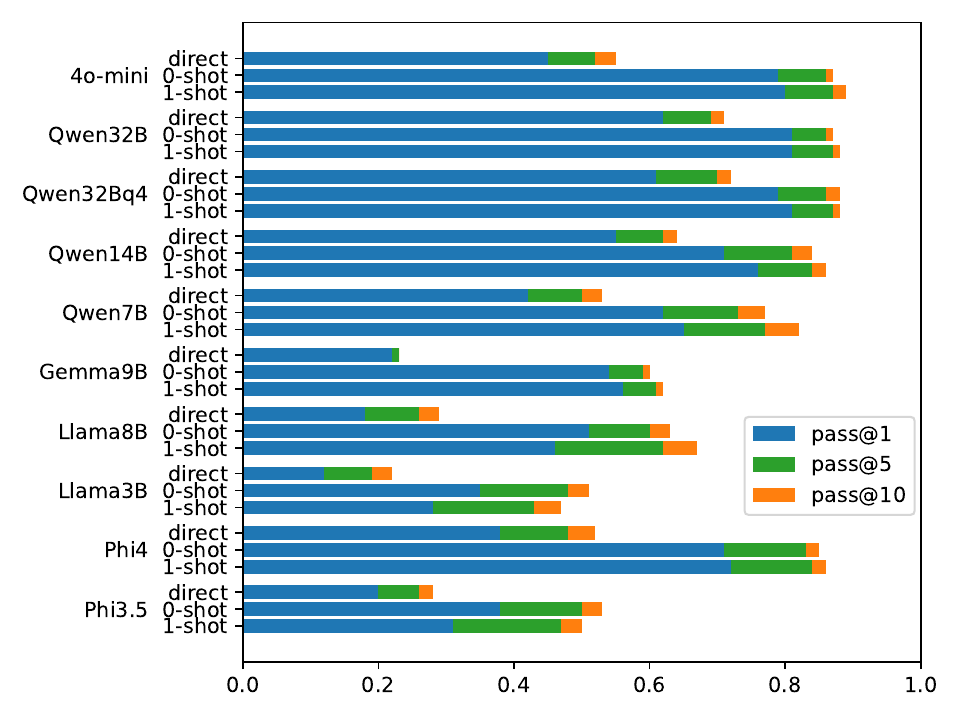} %
         \caption{To C++}
    \end{subfigure}
    \begin{subfigure}[b]{\textwidth}
         \centering
         \includegraphics[width=0.6\linewidth]{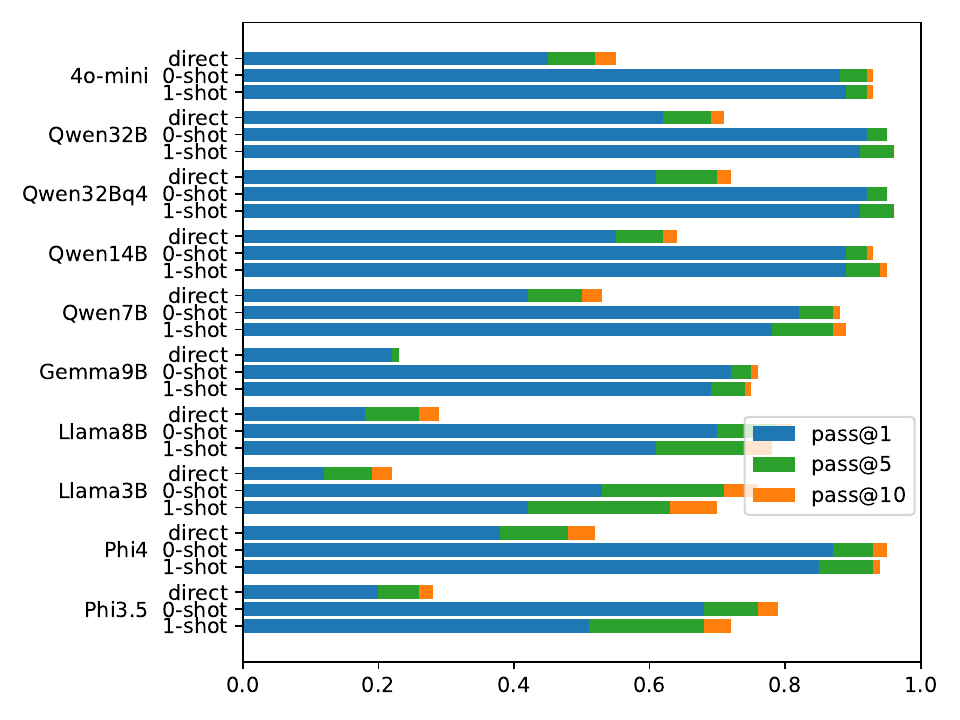}
         \caption{To Python}
    \end{subfigure}
    \begin{subfigure}[b]{\textwidth}
         \centering
         \includegraphics[width=0.6\linewidth]{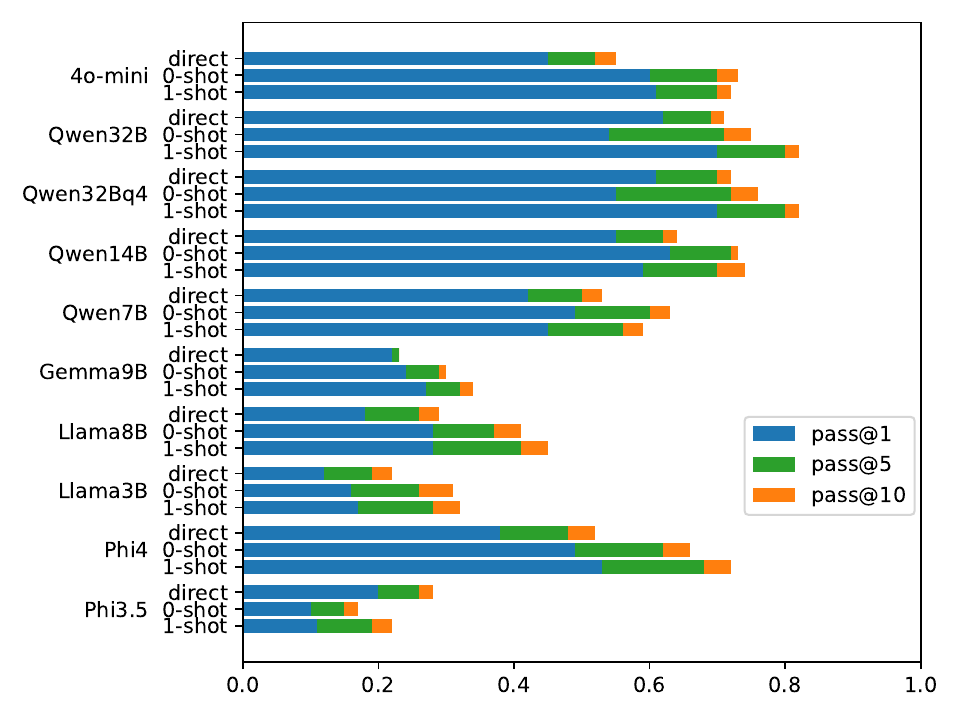}
         \caption{To Rust}
    \end{subfigure}
  \caption {Pass@k of code generation from pseudocode from Python to all languages, compared with direct generation from problems}
  \label{fig:compare-config-python}
\end{figure*}

\begin{figure*}[t]
\centering
    \begin{subfigure}[b]{\textwidth}
         \centering
         \includegraphics[width=0.6\linewidth]{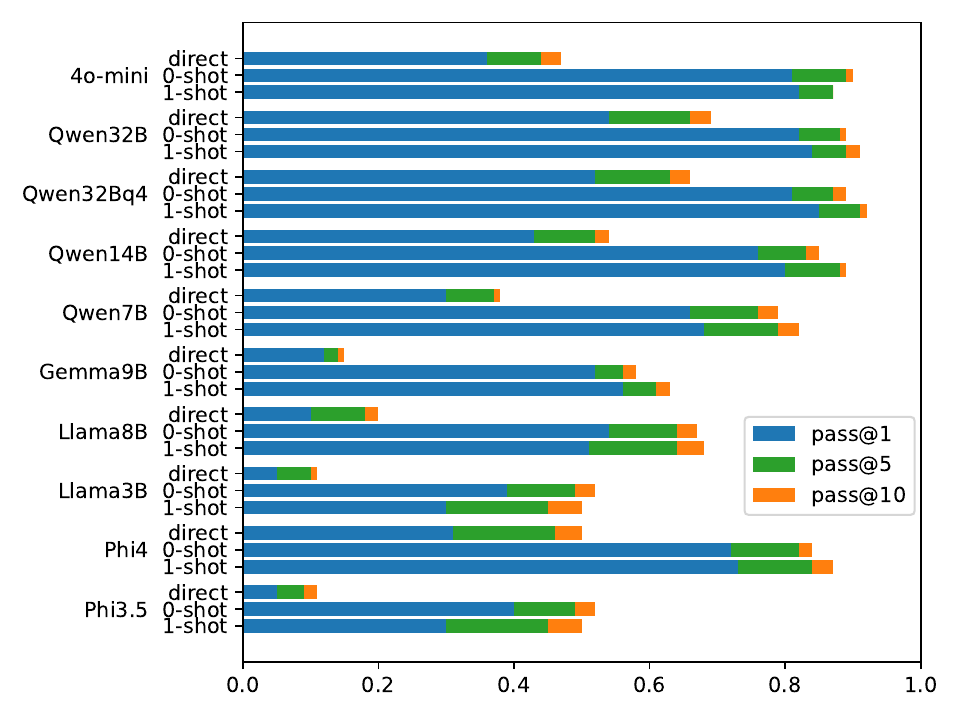} %
         \caption{To C++}
    \end{subfigure}
    \begin{subfigure}[b]{\textwidth}
         \centering
         \includegraphics[width=0.6\linewidth]{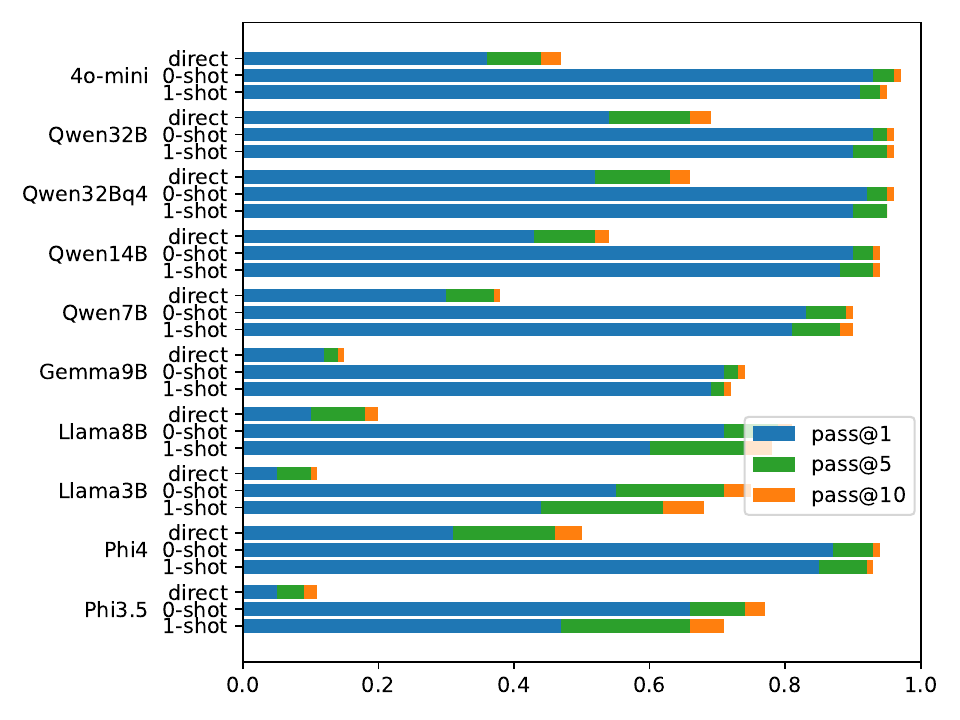}
         \caption{To Python}
    \end{subfigure}
    \begin{subfigure}[b]{\textwidth}
         \centering
         \includegraphics[width=0.6\linewidth]{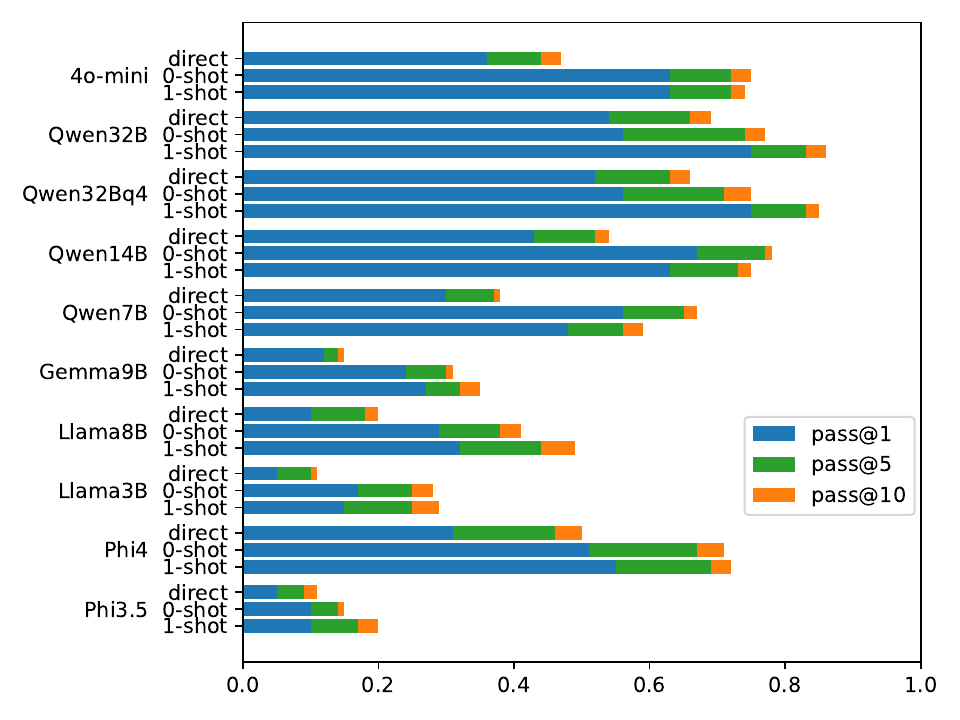}
         \caption{To Rust}
    \end{subfigure}
  \caption {Pass@k of code generation from pseudocode from Rust to all languages, compared with direct generation from problems}
  \label{fig:compare-config-rust}
\end{figure*}

\Cref{fig:compare-config-cpp}, \ref{fig:compare-config-python}, and \ref{fig:compare-config-rust} show the pass@k of code generation from pseudocode from C++, Python, and Rust, respectively, compared with the Pass@k of code generation from the problem.

\begin{figure*}
\centering
\footnotesize
\begin{lstlisting}
You are given an integer array nums and an integer k.
An integer h is called valid if all values in the array that are strictly greater than h are identical.
For example, if nums = [10, 8, 10, 8], a valid integer is h = 9 because all nums[i] > 9 are equal to 10, but 5 is not a valid integer.
You are allowed to perform the following operation on nums:

Select an integer h that is valid for the current values in nums.
For each index i where nums[i] > h, set nums[i] to h.

Return the minimum number of operations required to make every element in nums equal to k. If it is impossible to make all elements equal to k, return -1.
 
Example 1:

Input: nums = [5,2,5,4,5], k = 2
Output: 2
Explanation:
The operations can be performed in order using valid integers 4 and then 2.

Example 2:

Input: nums = [2,1,2], k = 2
Output: -1
Explanation:
It is impossible to make all the values equal to 2.

Example 3:

Input: nums = [9,7,5,3], k = 1
Output: 4
Explanation:
The operations can be performed using valid integers in the order 7, 5, 3, and 1.

 
Constraints:

1 <= nums.length <= 100 
1 <= nums[i] <= 100
1 <= k <= 100
\end{lstlisting}
\captionof{lstlisting}{Full problem in the motivating example}\label{lst:problem-motiv}
\end{figure*}

\begin{figure*}
\centering
\begin{lstlisting}[breaklines=true, language=C++, frame=shadowbox, numbers=left,]
class Solution {
public:
    int minOperations(vector<int>& nums, int k) {
        int mn = *min_element(nums.begin(), nums.end()); 
        if (mn < k) {
            return -1; 
        }
        unordered_map<int,int> mp; 
        for (auto &it: nums) {
            mp[it] = 1; 
        }
        int ans = mp.size(); 
        if (mp[k]) {
            ans--; 
        }
        return ans; 
    }
};
\end{lstlisting}
\captionof{lstlisting}{User-submitted C++ solution to \Cref{lst:problem-motiv}}
\label{lst:cpp-motiv}
\end{figure*}

\begin{figure*}
\centering
\begin{lstlisting}[breaklines=true, language=C++, frame=shadowbox, numbers=left,]
class Solution {
public:
    int nonSpecialCount(int l, int r) {
        // Calculate the limit up to which we need to find prime numbers
        int lim = (int)(sqrt(r));

        // Create a boolean array to mark primes up to lim using Sieve of Eratosthenes
        vector<bool> v(lim + 1, true);
        v[0] = v[1] = false; // 0 and 1 are not prime numbers

        // Sieve of Eratosthenes to mark non-prime numbers
        for (int i = 2; i * i <= lim; i++) {
            if (v[i]) {
                for (int j = i * i; j <= lim; j += i) {
                    v[j] = false;
                }
            }
        }

        // Count special numbers in the range [l, r]
        int cnt = 0;
        for (int i = 2; i <= lim; i++) {
            if (v[i]) {
                int square = i * i;
                if (square >= l && square <= r) {
                    cnt++;
                }
            }
        }

        // Total numbers in the range [l, r]
        int totalCount = r - l + 1;

        // Calculate non-special numbers
        return totalCount - cnt;
    }
};
\end{lstlisting}
\captionof{lstlisting}{A C++ solution that can be simplified }
\label{lst:cpp-simp}
\end{figure*}

\begin{figure*}
\centering
\begin{lstlisting}[breaklines=true, frame=shadowbox, numbers=left,]
function nonSpecialCount(l, r):
    lim = floor of sqrt(r)
    generate sieve for primes up to lim using Sieve of Eratosthenes
    cnt = count of primes i in 2..lim where i^2 is in [l, r]
    return (r - l + 1) - cnt
\end{lstlisting}
\captionof{lstlisting}{Pseudocode from \Cref{lst:cpp-simp}}
\label{lst:pseudo-simp}
\end{figure*}

\begin{figure*}
\centering
\begin{lstlisting}[breaklines=true, language=Rust, frame=shadowbox, numbers=left,]
impl Solution {
  pub fn maximum_points(mut arr: Vec<i32>, mut curr: i32) -> i64 {
    arr.sort_unstable();
    let mut res = 0;
    let mut pos = arr.len() - 1;

    while pos <= arr.len() && arr[0] <= curr {
      res += (curr / arr[0]) as i64;
      curr %
      curr += arr[pos];
      pos -= 1;
    }

    return res;
  }
}
\end{lstlisting}
\captionof{lstlisting}{Example user-submitted Rust code with subtraction underflow problem (line 11)}
\label{lst:rust-underflow}
\end{figure*}

\begin{figure*}
\centering
\begin{lstlisting}[breaklines=true, frame=shadowbox, numbers=left,]
function maximum_points(arr, curr)
    sort arr in ascending order
    res = 0
    pos = arr.len() - 1

    while pos >= 0 and arr[0] <= curr:
        res += curr // arr[0]
        curr = curr %
        curr += arr[pos]
        pos -= 1

    return res
\end{lstlisting}
\captionof{lstlisting}{Pseudocode converted from \Cref{lst:rust-underflow}}
\label{lst:pseudo-rust-underflow}
\end{figure*}

\begin{figure*}
\centering
\begin{lstlisting}[breaklines=true, language=Rust, frame=shadowbox, numbers=left,]
impl Solution {
    pub fn maximum_points(enemy_energies: Vec<i32>, current_energy: i32) -> i64 {
        let mut arr = enemy_energies.clone();
        arr.sort();
        let mut res: i64 = 0;
        let mut pos = arr.len() as isize - 1;
        let mut curr = current_energy;

        while pos >= 0 && arr[0] <= curr {
            res += (curr / arr[0]) as i64;
            curr = curr %
            if pos >= 0 {
                curr += arr[pos as usize];
            }
            pos -= 1;
        }

        res
    }
}
\end{lstlisting}
\captionof{lstlisting}{Correct Rust code implemented from \Cref{lst:pseudo-rust-underflow}}
\label{lst:rust-fix-underflow}
\end{figure*}

\begin{figure*}
\centering
\begin{lstlisting}[breaklines=true, language=Python, frame=shadowbox, numbers=left,]
class Solution:
    def maximumSubarraySum(self, nums: List[int], k: int) -> int:
        max_sum, prefix_sum, val_to_min_prefix_sum = -inf, 0, defaultdict(lambda: inf)
        for i, num in enumerate(nums):
            if val_to_min_prefix_sum[num] > prefix_sum:
                val_to_min_prefix_sum[num] = prefix_sum
            prefix_sum += num
            max_sum = max(max_sum, prefix_sum - val_to_min_prefix_sum[num + k], prefix_sum - val_to_min_prefix_sum[num - k])
        return max_sum if max_sum > -inf else 0 
\end{lstlisting}
\captionof{lstlisting}{Python code whose converted pseudocode has negative impact}
\label{lst:python-acc}
\end{figure*}

\begin{figure*}
\centering
\begin{lstlisting}[breaklines=true, frame=shadowbox, numbers=left,]
function circularGameLosers(n, k):
    s = {1, 2, ..., n}
    for each mul in cumulative sums of integers from 0 to n-1:
        position = (k * mul) mod n + 1
        if position not in s:
            break
        remove position from s
    return s as a list
\end{lstlisting}
\captionof{lstlisting}{Pseudocode converted from \Cref{lst:python-acc}}
\label{lst:pseudo-python-acc}
\end{figure*}

\begin{figure*}
\centering
\begin{lstlisting}[breaklines=true, language=Python, frame=shadowbox, numbers=left,]
class Solution:
    def circularGameLosers(self, n: int, k: int) -> List[int]:
        players = set(range(1, n + 1))
        mul = 0
        while True:
            position = (k * mul) %
            if position not in players:
                break
            players.remove(position)
            mul += 1
        return sorted(players)
\end{lstlisting}
\captionof{lstlisting}{Wrong Python code generated from \Cref{lst:pseudo-python-acc} }
\label{lst:python-acc-wrong-gen}
\end{figure*}

\begin{figure*}
\centering
\begin{lstlisting}[breaklines=true, language=Python, frame=shadowbox, numbers=left,]
class Solution:
    def maximumSubarraySum(self, nums: List[int], k: int) -> int:
        max_sum, prefix_sum, val_to_min_prefix_sum = -inf, 0, defaultdict(lambda: inf)
        for i, num in enumerate(nums):
            if val_to_min_prefix_sum[num] > prefix_sum:
                val_to_min_prefix_sum[num] = prefix_sum
            prefix_sum += num
            max_sum = max(max_sum, prefix_sum - val_to_min_prefix_sum[num + k], prefix_sum - val_to_min_prefix_sum[num - k])
        return max_sum if max_sum > -inf else 0 
\end{lstlisting}
\captionof{lstlisting}{Python code being wrongly converted into pseudocode}
\label{lst:python-being-wrongly-pseudogen}
\end{figure*}

\begin{figure*}
\centering
\begin{lstlisting}[breaklines=true, frame=shadowbox, numbers=left,]
function maximumSubarraySum(nums, k):
    max_sum = -infinity
    prefix_sum = 0
    val_to_min_prefix = empty map  // default value is infinity
    for each num in nums:
        if val_to_min_prefix[num] > prefix_sum:
            val_to_min_prefix[num] = prefix_sum
        prefix_sum += num
        candidate1 = prefix_sum - val_to_min_prefix[num + k]
        candidate2 = prefix_sum - val_to_min_prefix[num - k]
        max_sum = max(max_sum, candidate1, candidate2)
    return max(max_sum, 0)
\end{lstlisting}
\captionof{lstlisting}{Pseudocode converted from \Cref{lst:python-being-wrongly-pseudogen} with errors (line 12) }
\label{lst:pseudo-wrong-condition}
\end{figure*}

\begin{figure*}
\centering
\footnotesize
\begin{lstlisting}
I am a Python programmer.
Please help me convert Python code into a semantic-preserving and concise pseudocode.
Instead of translating line by line, you should simplify the pseudocode as much as possible and also readable.
Below are specific rules:

1. Use indents to represent control structures.
```
if a == b:
    c += 1
```

2. The pseudocode should not be tied to a specific programming language and should not contain any language-specific stuffs such as `yield` in Python.

3. The pseudocode does not need to preserve concrete type info: (a) The concrete names such as `vector` and `i64` should not appear. Usually, general names such as array/list and int are enough for describing algorithms. (b) Do not involve type casting.

4. You should omit the implementation of common algorithms/data structures/operations.

For example, the customized binary search subroutine
```
def search_square_geq(nums, val):
    left = 0
    right = len(nums) - 1
    while left < right:
        mid = left + (right - left) // 2
        if nums[mid]**2 < val:
            left = mid + 1
        else:
            right = mid
    return left
    
target = search_square_geq(xs, 9)
```
can be simplified as
```
target = binary search for the index i such that xs[i] * xs[i] >= 9
```

5. You can use natural language to simplify code, in particular loops. For example,
```
for x in xs:
    if x == 233:
        flag = true
```
can be simplified as `flag = whether 233 exists in xs`

6. Do not use natural language if that is verbose. For example, `let n be the size of list_a` is less compact and readable than `n = list_a.size()`

7. A function definition should be formatted like `function max(a, b)`. Functions can be nested and can use variables in the outer scope.

Finally, recall that the principles are **semantic-preserving** and **concise and readable**.
Do not change the name of the given function.
You can iterate the writing of pseudocode to ensure it follows the above rules.
Wrap only the final version with code blocks (```) in the response.

Below is the Python code to convert into pseudocode.
{code}
\end{lstlisting}
\captionof{lstlisting}{Prompt (a single user query) to generate pseudocode from DeepSeek-R1 }\label{lst:pseudogen-prompt}
\end{figure*}

\begin{figure*}
\centering
\footnotesize
\begin{lstlisting}
===System===
You are a proficient {lang} programmer and familiar with various algorithms.
Your task is to implement a {lang} code given a pseudocode illustrating an algorithm and a {lang} signature.
If some subroutines are provided in the standard library, you can use them.
For subroutines that are not provided in the standard library, you need to implement them.

===User===
```
{pseudocode}
```

Please implement the above pseudocode with the following template.
```
{signature}
```
\end{lstlisting}
\captionof{lstlisting}{Zero-shot prompt to generate code from pseudocode  }\label{lst:query-prompt-zero}
\end{figure*}

\begin{figure*}
\centering
\footnotesize
\begin{lstlisting}
===System===
You are a proficient Python programmer and familiar with various algorithms.
Your task is to implement a Python code given a pseudocode illustrating an algorithm and a Python signature.
If some subroutines are provided in the standard library, you can use them.
For subroutines that are not provided in the standard library, you need to implement them.
Below is a simple example.

===User===
```pseudocode
Input:
    nums is a list of positive integers;
    k is a positive integer
Output:
    return x such that x is the smallest in nums and x * x > k
function findTarget(nums, k):
    sorted = sort nums in ascending order
    res = binary search on sorted for the leftmost value res such that res * res > k
    return res
```

Please implement the function with the following template.
```python
class Solution:
    def findTarget(self, nums: List[int], k: int) -> int:
        
```

===Assistant===
The pseudocode involves two subroutines, sort and binary search.
Sort can be implemented using the standard libray, and binary search needs to be implemented by ourselves.
Finally, the Python implementation is as follows.

```python
class Solution:
    def findTarget(self, nums: List[int], k: int) -> int:
        nums = sorted(nums)
        left = 0
        right = len(nums) - 1
        while left < right:
            mid = left + (right - left) // 2
            if nums[mid] * nums[mid] <= k:
                left = mid + 1
            else:
                right = mid
        return nums[left]
```

===User===
```
{pseudocode}
```

Please implement the function with the following template.
```
{signature}
```
\end{lstlisting}
\captionof{lstlisting}{One-shot prompt to generate code from pseudocode  }\label{lst:query-prompt-one}
\end{figure*}

\end{document}